\documentclass[aps,prl,reprint, superscriptaddress]{revtex4-1}
\usepackage{graphicx}
\usepackage{caption}
\usepackage{subcaption}
\usepackage{amsmath}
\usepackage{amsfonts, amsmath, amssymb, bm, graphicx}
\graphicspath{{pictures/}}
\DeclareGraphicsExtensions{.pdf,.png,.jpg,.eps}

\usepackage{textcomp}

\def\<{\left\langle}
\def\>{\right\rangle}

\def\const{{\rm const}}

\begin{document}

\title{Ultra-fast artificial neuron: generation of picosecond-duration spikes in a current-driven antiferromagnetic auto-oscillator}

\author{Roman~Khymyn}
\thanks{R. Khymyn and I. Lisenkov contributed equally to this work}
\affiliation{Department of Physics, University of Gothenburg, 41296 Gothenburg, Sweden}

\author{Ivan Lisenkov}
\thanks{R. Khymyn and I. Lisenkov contributed equally to this work}
\affiliation{Department of Electrical Engineering and Computer Science, Oregon State University, Corvallis, OR 97331, USA}
\affiliation{Institute of Radioengineering and Electronics of RAS, Moscow, Russia}

\author{Jamie Voorheis}
\affiliation{Department of Physics, Oakland University, Rochester, MI 48309, USA}

\author{Olga Sulymenko}
\affiliation{Faculty of Radiophysics, Electronics and Computer Systems, Taras Shevchenko National University of Kyiv, Kyiv, 01601, Ukraine}

\author{Oleksandr Prokopenko}
\affiliation{Faculty of Radiophysics, Electronics and Computer Systems, Taras Shevchenko National University of Kyiv, Kyiv, 01601, Ukraine}

\author{Vasil Tiberkevich}
\affiliation{Department of Physics, Oakland University, Rochester, MI 48309, USA}

\author{Johan ~Akerman}
\affiliation{Department of Physics, University of Gothenburg, 41296 Gothenburg, Sweden}

\author{Andrei Slavin}
\affiliation{Department of Physics, Oakland University, Rochester, MI 48309, USA}


\begin{abstract}
We demonstrate analytically and numerically, that a thin film of an antiferromagnetic (AFM) material, having biaxial magnetic anisotropy and being driven by an external spin-transfer torque signal, can be used for the generation of ultra-short "Dirac-delta-like" spikes. The duration of the generated spikes is several picoseconds for typical AFM materials, and is determined by the in-plane magnetic anisotropy and the effective damping of the AFM material. The generated output signal can consist of a single spike or a discrete group of spikes ("bursting"), which depends on the repetition (clock) rate, amplitude  and shape of the external control signal. The spike generation occurs only when the amplitude of the control signal exceeds a certain threshold, similar to the action of a biological neuron in response to an external stimulus. The "threshold" behavior of the proposed AFM spike generator makes possible its application not only in the traditional microwave signal processing, but also in the future neuromorphic signal processing circuits working at clock frequencies of tens of gigahertz .
\end{abstract}

\maketitle


Generators of short in the time domain ``Dirac-delta-like'' pulses are widely used in modern electronics and optics. The most obvious applications of such generators are for the formation of trigger sequences, pulse-density modulation (PDM) of signals, and other signal processing purposes. In PDM the amplitude of an analog input signal is encoded by the relative repetition rate of the generated pulses. A train of ``Dirac-delta-like'' pulses with a constant repetition rate forms a Fourier image of equidistant sharp peaks in the frequency domain, which is known as a frequency ``comb''. Generators of frequency-``combs'' are used for dense frequency-division multiplexing (DFDM) in electronics, and dense linewidth-division multiplexing (DLDM) in optics to exploit the full bandwidth of the data transmission lines. In this approach, the generators of the carrier frequencies can be locked to a corresponding frequency of the ``comb''. Thus, one of the key characteristics of signal processing devices using the ``comb'' generators is the bandwidth of the frequency-``comb'' generator, which is limited by the duration of a single pulse.

A similar type of pulse-encoded signals is used in nervous systems of biological objects, where response of a neuron to an input stimulus is a single spike, or a train of spikes with a certain sequence frequency, which is called an action potential in the cell biology~\cite{CellBiology}. Therefore, the modern concepts of neuromorphic computing and signal processing include spike generators as a mandatory element of their architecture \cite{indiveri2011neuromorphic, aoyagi2001bursting}. Another peculiarity of a nervous system is the neuron's threshold behavior, in a sense, that a neuron generates a response \emph{only} when the input stimulus is above a certain critical value (threshold). This  nonlinear response is also a key feature for the operation of artificial neuromorphic devices.

Electronic frequency-"comb" generators, usually, employ the modern CMOS technology, and can have a compact design, but operate at relatively low frequencies, and, therefore, have a relatively low frequency bandwidth ($<50\text{GHz}$) \cite{tomita20138}. The optical comb generators offering microwave frequency spacing, based on the phase modulation in Fabry-Perot cavities \cite{saitoh1995waveguide}, multi-frequency lasers \cite{park199624}, Brillouin-enhanced fiber lasers \cite{yamashita1998bidirectional} and phase modulation within an amplified fiber loop \cite{bennett19991} can have a substantially wider frequency span ($>100\text{GHz}$), but are rather complex devices, incompatible with the existing on-chip technology.

Spin-torque nano-ocillators (STNO) and spin-Hall oscillators (SHO) based on ferromagnetic (FM) materials \cite{Kiselev2003, Demidov2012, Demidov2016, chen2016spin, mohseni2013spin} are of a high interest for modern spintronics as tunable nano-scale generators of microwave signals, and, in principle, can be used as pulse generators, but their typical response time is determined by the frequency of the ferromagnetic resonance, and is limited to hundreds of picoseconds by the practically achievable magnitudes of the local bias magnetic field.

Recently, however, it has been proposed to use AFM materials as active layers of SHOs due to their ability to operate at higher frequencies, up to the THz range \cite{bib:Gomonay:2014, bib:Cheng:2016, cheng2014spin, bib:Liu:2016, khymyn2017antiferromagnetic, sulymenko2017terahertz}. In an AFM-based SHO, the spin current $j_s^{in}$ created by the spin-Hall effect (SHE) in the adjacent heavy metal layer traversed by a direct electric current, induces a torque on the Neel vector $\mathbf{l}$ of the AFM. If the spin polarization $\mathbf{p}$ of the driving current in AFM is perpendicular to the equilibrium orientation of the Neel vector $\mathbf{l}_0$,  the Neel vector starts to rotate in the plane perpendicular to the vector $\mathbf{p}$ \cite{bib:Gomonay:2014, khymyn2017antiferromagnetic, sulymenko2017terahertz}.

The extraction of an ac signal by inverse SHE (ISHE) in the adjacent layer of a heavy metal is, however, non-trivial, because it requires the motion of the Neel vector that is non-uniform in time. Several approaches were proposed to solve this problem \cite{bib:Cheng:2016, khymyn2017antiferromagnetic, sulymenko2017terahertz}. For instance, it was shown in Ref.~\cite{khymyn2017antiferromagnetic}, that a  non-uniform rotation of the Neel vector can be achieved in AFM materials with bi-axial type of anisotropy (e.g., NiO), where additional in-plane anisotropy creates an effective potential profile for the rotating Neel vector. The output signal of the AFM generators based on this mechanism, is however, a  simple harmonic (sinusoidal) oscillation.

\begin{figure}
	\includegraphics[width=0.65\linewidth]{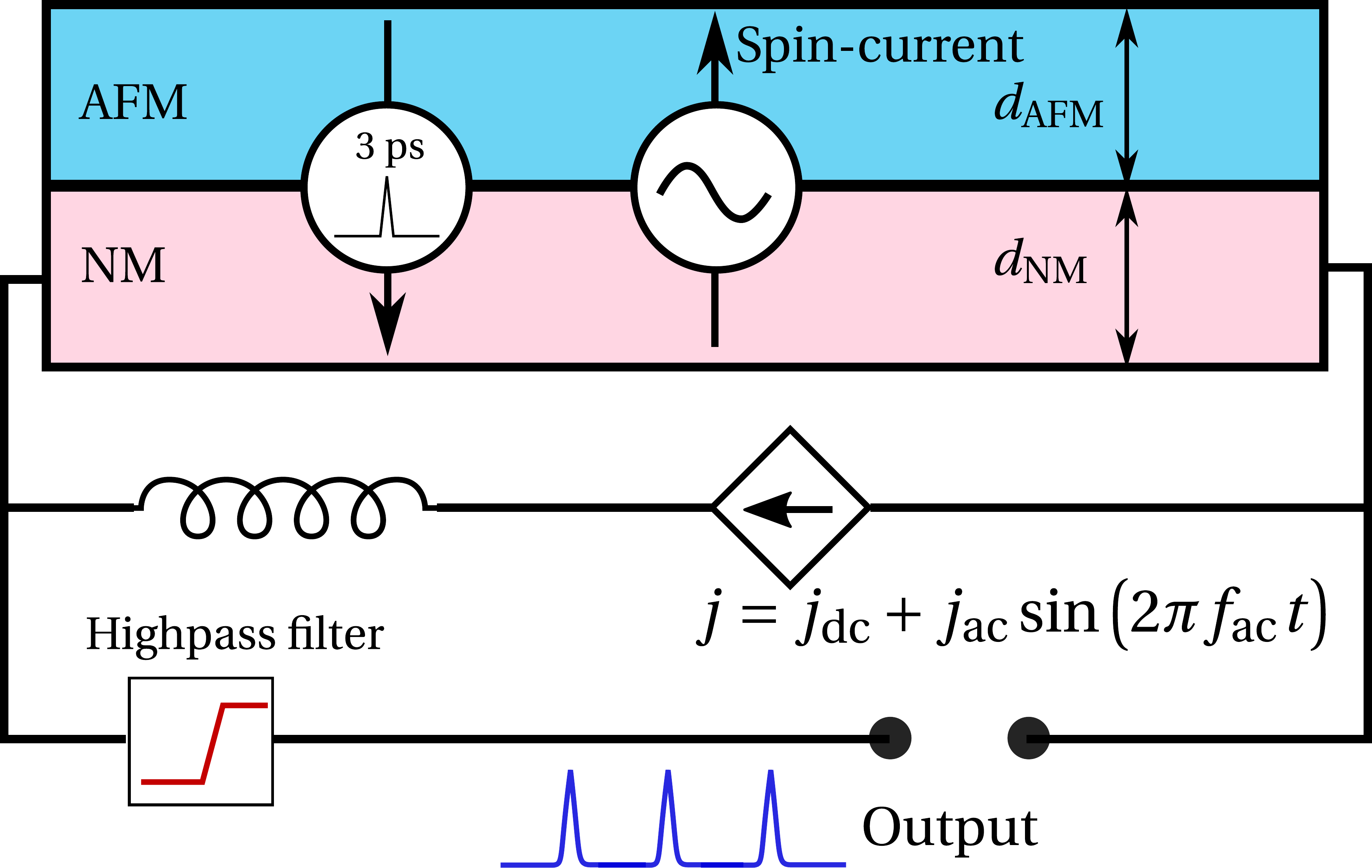}
	\caption{
	Principal operational scheme of a spintronic generator of ultra short pulses (spikes). A thin antiferromagnetic (AFM) layer is covered by a layer of a normal metal (NM). The NM layer is driven by an ac-modulated electrical current. The output signal is received through a high-pass filter allowing only the frequencies higher than the modulation frequency to get out.
	}
	\label{fig:scheme}
\end{figure}

Here, we propose a design of an AFM-based  spin-Hall auto-oscillator, capable of generating controlled sequences of ultrashort pulses with a typical pulse duration of a few picoseconds. The proposed generator is based on a layered structure consisting of a current-driven layer of a normal metal (NM) with a strong spin orbit-coupling, and an antiferromagnetic (AFM) layer with a biaxial magnetic anisotropy. Spin current ($j_s^{in}$) created by the spin-Hall effect (SHE) in the NM and flowing into a thin AFM film creates a torque on the sublattice magnetizations of the AFM, which leads to a rapid switch of their orientation. This switch, in its turn, creates a short pulse of the spin current flowing back to the NM layer ($j_s^{out}$), where it can be converted into an electrical signal by the inverse spin-Hall effect (ISHE), see Fig.~\ref{fig:scheme}.

The minimum duration of a short pulse is limited by a characteristical time of the current-induced AFM sublattice reorientation process. For a relative strong magnetic damping, this reorientation time is proportional to the effective Gilbert damping constant and inversely proportional to the magnitude of the in-plane magnetic anisotropy field in the AFM material (see Eq.~\eqref{eq:duration} below). Since the effective Gilbert damping constant in sandwiched AFM/NM structures is mostly determined by the spin-pumping~\cite{Tserkovnyak2002}, it can be controlled, and the achievable reorientation time can be of the order of several picoseconds for typical AFM/NM bilayers, such as nickel oxide (NiO) and Pt. The pulse repetition rate in the proposed generator of ultrashort pulses can be easily controlled  by the frequency of the input electrical current, and can reach values exceeding $100\text{GHz}$. The short duration of a single generated pulse, also, defines a wide frequency span of the generated frequency comb -- $200\text{GHz}$ at $-10\text{dB}$ for NiO.

The generation of short spikes in the above proposed AFM-based pulse generator happens to be a \emph{threshold process} with respect to the amplitude of the input driving electric current, which gives us  a possibility to consider such an oscillator as an~\emph{ultra-fast analogue of a  biological neuron}. Moreover, the complex behavior of a neuron, such as generation of  discrete groups of spikes (bursting \cite{aoyagi2001bursting}), can be also achieved  in the proposed AFM generator for certain parameters of the  driving  ac electric signal.


Below, we discuss the geometry of the AFM SHO, considered in Ref.~\cite{khymyn2017antiferromagnetic}. In that case a thin film of an AFM material has two magnetic sublattices with magnetizations $\mathbf{M}_1$ and $\mathbf{M}_2$, and biaxial magnetic anisotropy: the anisotropy easy plane is perpendicular to the axis $\mathbf{n}_h$, and its effective magnetic field is $H_h$, while the anisotropy easy axis is directed along the vector $\mathbf{n}_e$, and its effective magnetic  field  is $H_e$. We shall also assume, that the polarization of the spin current flowing into the AFM layer is directed along the hard axis of the AFM anisotropy $\mathbf{p} \parallel \mathbf{n}_h$.

The spin dynamics of the AFM with bi-axial anisotropy under the action of an external spin torque is described by the set of two coupled Landau-Lifshitz equations for the  magnetization of each of the AFM sublattices $\textbf{M}_{1}$ and $\textbf{M}_{2}$. However, in the low frequency limit (i. e. when the frequency of oscillations is much less than the exchange frequency $\approx30\text{THz}$) for the considered geometry of an AFM SHO this set of equations can be reduced to the single equation describing the dynamics of the rotation angle $\phi$ of the Neel vector $\mathbf{l}=(\mathbf{M}_1-\mathbf{M}_2)/2M_s$ in the easy plane of the AFM \cite{khymyn2017antiferromagnetic}. To make everything well-defined, we assume that the angle $\phi$ is measured from the easy axis $\mathbf{n}_e$ ($\mathbf{l} \cdot \mathbf{n}_e=\cos \phi$). In terms of the angle $\phi$ the equation for the Neel vector dynamics has the following form \cite{khymyn2017antiferromagnetic}:

\begin{equation}
\frac{1}{\omega_{ex}} \ddot \phi + \alpha_{\rm eff} \dot \phi+\frac{\omega_e}{2} \sin 2 \phi = \sigma j_e (t),
\label{eq:pendulum}
\end{equation}
where $\omega_{ex}=\gamma H_{ex}$ is the exchange frequency and $\omega_e=\gamma H_e$. The term in the right-hand-side part of Eq. (\ref{eq:pendulum}) describes the spin torque created by the electrical current $j_e(t)$ flowing in NM layer (expressed in the frequency units)\cite{bib:Nakayama:2012, khymyn2017antiferromagnetic}. The effective damping parameter $\alpha_{\rm eff}=\alpha_0+\alpha_{SP}$ includes both the intrinsic Gilbert damping constant $\alpha_0$ and the additional magnetic losses due to the spin pumping from the adjacent layer of the normal metal $\alpha_{SP}$. The losses due to the spin pumping depend on the  thickness $d_{AFM}$ of the AFM layer ($\alpha_{SP} \sim 1/d_{AFM}$), and, therefore, can be adjusted in a certain range by a proper design of the geometric parameters of the AFM SHO.

\begin{figure}
	\centering
	\includegraphics[width=0.95\linewidth]{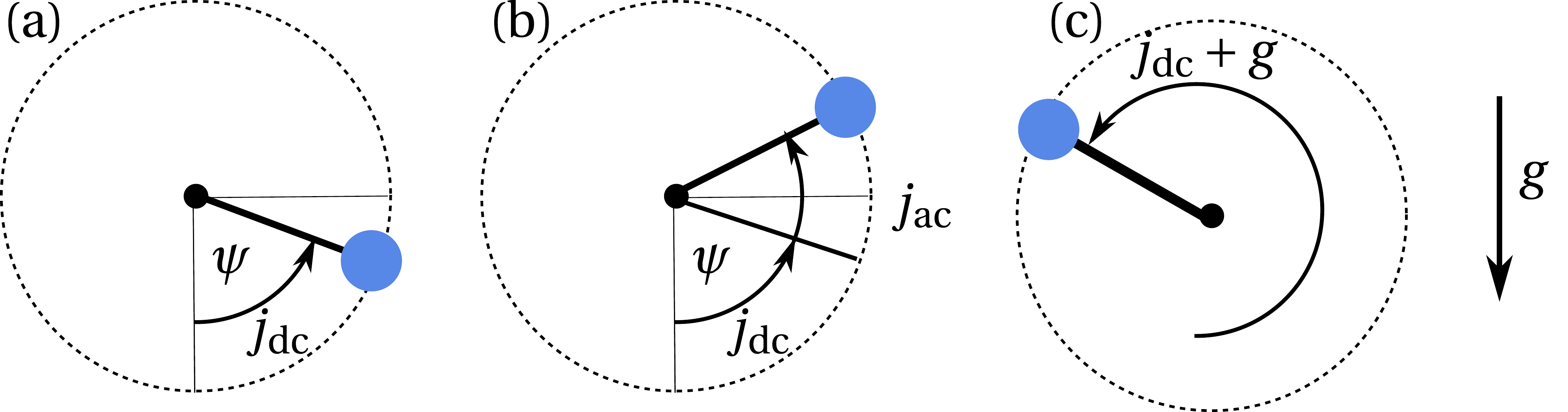}
	\caption{
		The dynamics of a physical pendulum as an analogue to the dynamics of the AFM generator. Angle $\psi$ of the pendulum corresponds to $\phi/2$ angle of the Neel vector in the AFM: a) ``below threshold'' torque $j_\text{dc}$ lifts the pendulum from the ground state; b) additional controllable torque $j_\text{ac}$ transfers the oscillator through the threshold;
 c) fast motion of the pendulum, which corresponds to the AFM sublattice reorientation.}
	\label{fig:pendulum}
\end{figure}

The first term in Eq. (\ref{eq:pendulum}) describes the inertial properties of an AFM SHO. However, when the damping is high ($2\alpha_{\text{eff}}>\pi \sigma j_{e}/ \sqrt{\omega_e\omega_{ex}}$) this term can be neglected in a qualitative analysis. It is really remarkable, that the spin dynamics of an AFM oscillator described by Eq. (\ref{eq:pendulum}) is mathematically analogous to the dynamics of an~\emph{over-damped} physical pendulum in a gravitational potential under the action of an external torque $\sigma j_e (t)$. In this analogy the   the magnetic anisotropy in the AFM layer plays the role of a gravitational field $g$, Gilbert damping plays the role of a friction, and the inverse of the exchange frequency plays the role of the  pendulum inertial mass. Since the gravity is a directional field and the magnetic anisotropy is bi-directional, to make the full analogy we have to replace the angle of the Neel vector $\phi$ with the angle of the pendulum $\psi$ as $2\phi = \psi$, see Fig.~\ref{fig:pendulum}(a). In the absence of the external torque $j_e=0$ the "AFM pendulum" is in a ground state, which defines the energy minimum $\psi = \phi=0$. A small, steady in time driving torque lifts the ``pendulum'' to a tilt angle $\psi_0=\arcsin(2\sigma j_{dc}/\omega_e)$. This happens when the dc current is flowing in the NM layer $j_e=j_{dc}=\const$). The maximum value of the tilt angle at a stationary state of the pendulum is $\psi_0^{max}=\pi/2$, because at this angle the returning torque from the "gravity" potential is maximized, and, if the torque overcomes this threshold $\sigma j_{dc}>\sigma j_{dc}^{th} = \omega_e/2$, the pendulum undergoes an infinite rotational motion~\cite{khymyn2017antiferromagnetic}.

Now, let us consider the dynamics of the pendulum when some additional torque $\sigma j_t$ is applied for a short period of time. In the initial situation the pendulum remains tiled with an angle $\phi_0$. After an additional torque $j_t$ is turned on, the angle $\phi$ starts to increase. If $j_{dc} + j_t > j_{dc}^{th}$, the pendulum overcomes the threshold angle $\psi>\pi/2$ .  At this point the torque created by the $j_{dc}$ alone is sufficient to continue the rotation of the pendulum, and, therefore, one could turn off the additional current $j_t$. Importantly, when $\psi>\pi$ the returning force, coming from the ``gravitational'' (or anisotropy) potential, is now assisting the torque $j_{dc}$, which results in a very fast acceleration of the pendulum in the region $\pi<\psi<2\pi$. If the damping is sufficiently large to stop the infinite rotation ($2\alpha_{eff}>\pi \sigma j_{dc}/ \sqrt{\omega_e\omega_{ex}}$), the pendulum will relax to a new stationary point $\phi_0+2\pi$, which corresponds to the switching of the magnetization sublattices $\mathbf{M}_1$ and $\mathbf{M}_2$ to an opposite direction. Note, that 
we are interested in the case of relatively small values of the control current $j_t$, that requires the bias current $j_{dc}$ to 
be close to the threshold value ($j_{dc}\simeq j_{dc}^{th}$, and the value of the effective damping in the AFM material to be rather large $4\alpha_{eff}>\pi \sqrt{\omega_e/\omega_{ex}}$.

\begin{figure}
	\centering
	\includegraphics[width=1.0\linewidth]{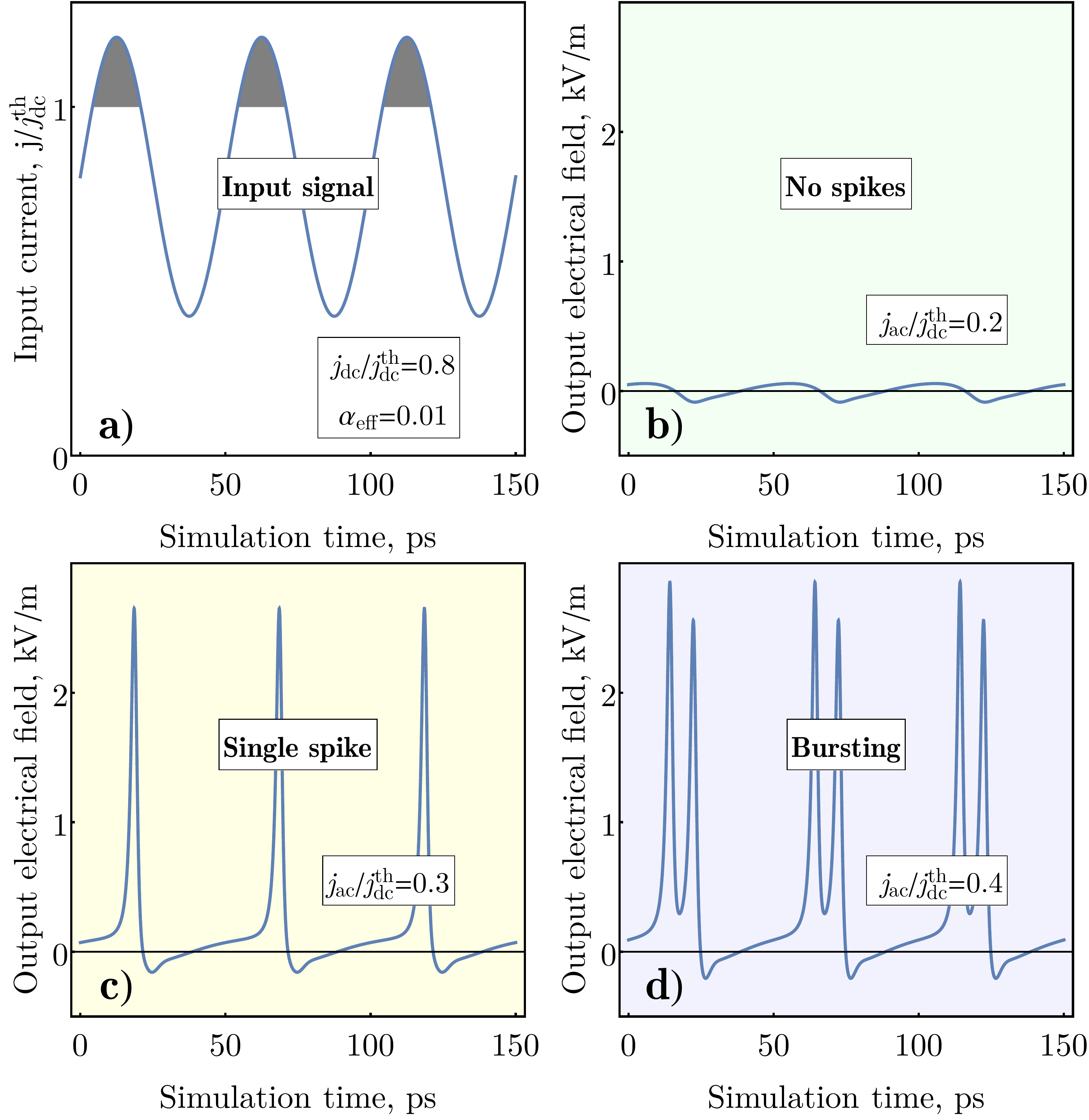}
	\caption{
  Numerical simulation of different regimes of spike generation in the AFM SHO : a)Shape of the combined (dc + ac) input driving current;  b) Regime of no generation ("low" value of the ac current $j_{ac}$); c)Generation of a single output spike during each period of the driving ac current ("moderate" values of the ac current $j_{ac}$ ); d)Generation of  a discrete group of spikes (bursting) during each period of the AC current ("high" values of the ac current $j_{ac}$).
	}
	\label{fig:regimes}
\end{figure}

\begin{figure*}
	\centering
	\includegraphics[width=0.8\textwidth]{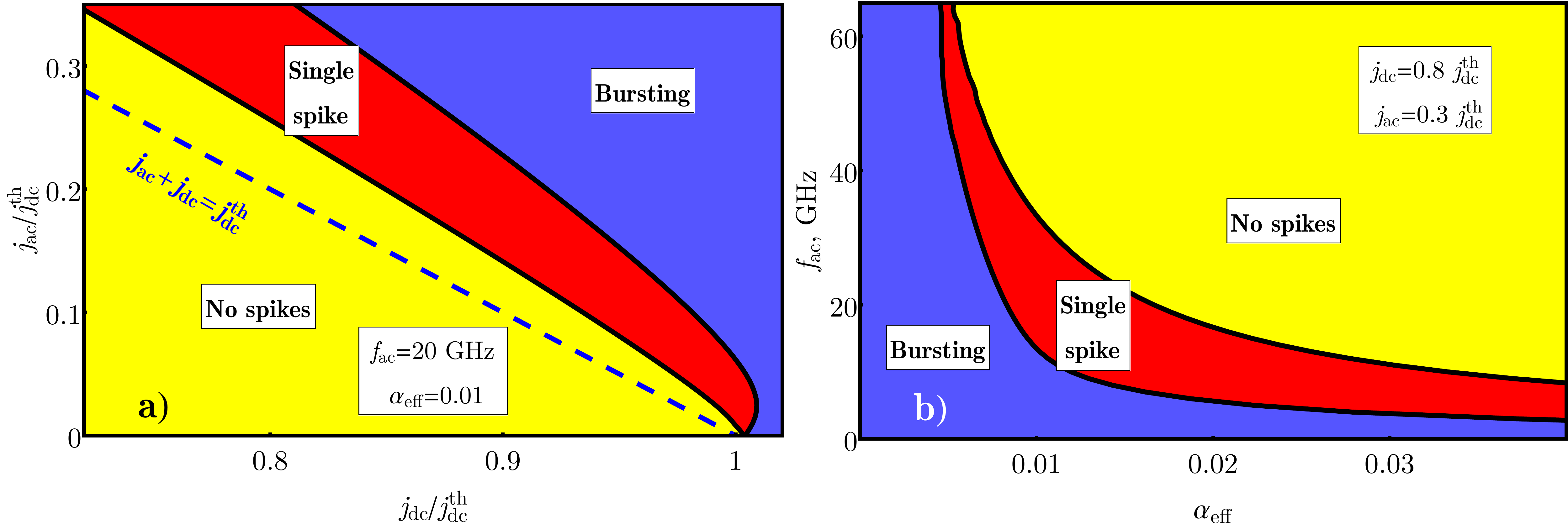}
	\caption{``Phase diagrams'' of the spike generation regimes for different values of the experimentally controllable parameters of the driving current $j_{dc}$, $j_{ac}$, $f_{ac}$ and the AFM material dissipation $\alpha_{eff}$: a)"Phase diagram" on the plane  $j_{ac}/ j_{dc}^{th}$ vs. $j_{dc}/ j_{dc}^{th}$  at the fixed value of the driving ac frequency $f_{ac}=$ 20 GHz and $\alpha_{eff} = 0.01$; b)"Phase diagram" on the plane $f_{ac}$ vs. $\alpha_{eff}$ at the fixed values of  $j_{dc}/ j_{dc}^{th}=0.8$  and  $j_{ac}/ j_{dc}^{th}=0.3$.
	}
	\label{fig:diagram}
\end{figure*}

\begin{figure}
	\centering
	\includegraphics[width=0.45\textwidth]{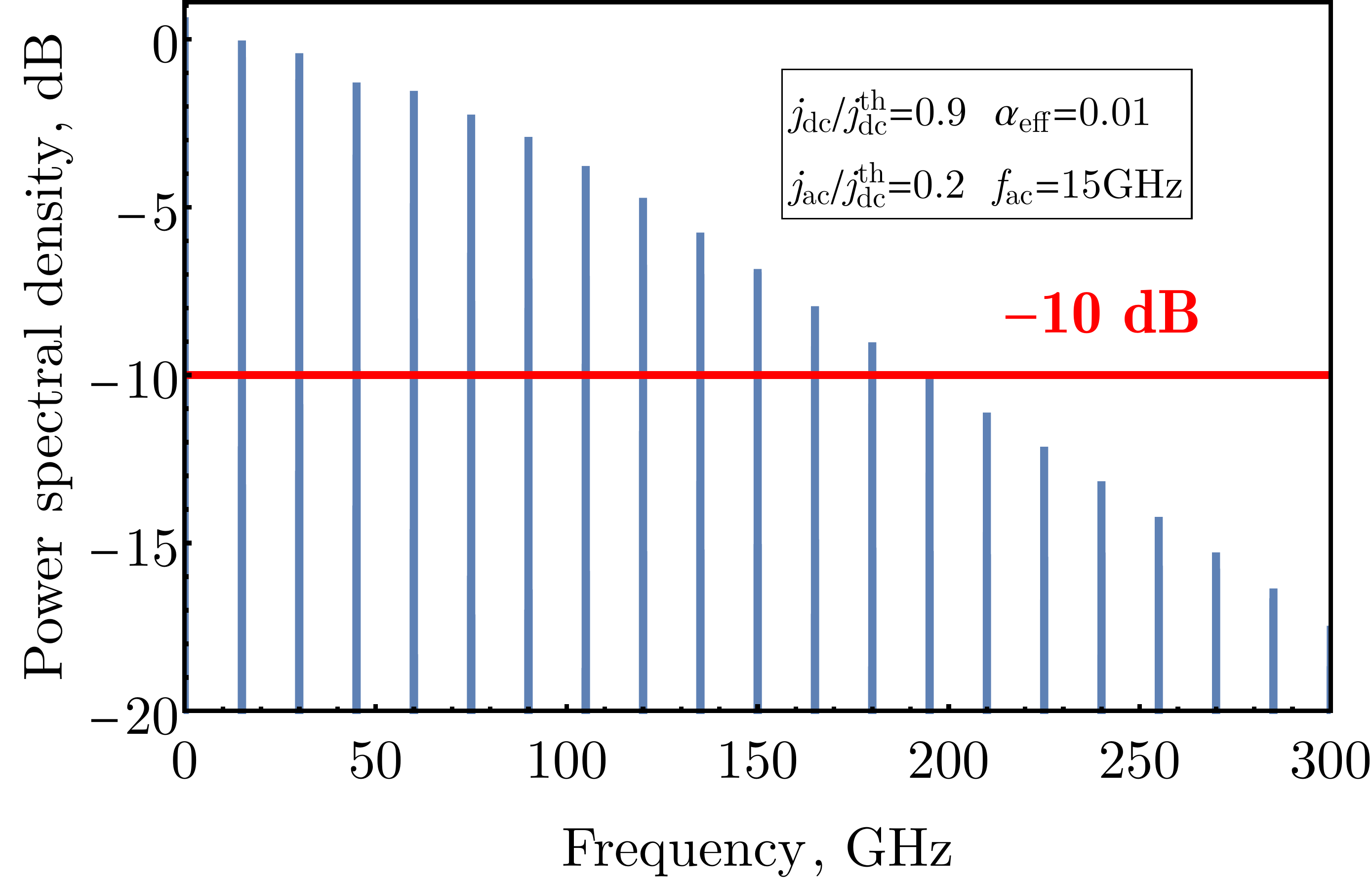}
	\caption{
		Numerically simulated frequency comb (spectrum of a periodic sequence of spikes) generated by an AFM SHO. Spectrum is normalized by the amplitude of the first harmonic at $f_{ac}=15\text{GHz}.$
	}
	\label{fig:FrequencyComb}
\end{figure}

The electrical field in the NM layer, produced by the back spin-pumping through the ISHE, is proportional to the angular velocity of the Neel vector: $E=\kappa \dot \phi$. The pendulum reaches maximum velocity $\dot \phi_{max}$ (which determines the spike amplitude)  at $\ddot \phi=0$, and for the above described scenario of the spike generation one gets from Eq. (\ref{eq:pendulum}): $\dot \phi_{max}=(\sigma j_{dc}+\omega_e/2)/{\alpha}$, or $\dot \phi_{max}=\sigma \omega_e/{\alpha}$ when $j_{dc}$ is close to the threshold value ($j_{dc}\simeq j_{dc}^{th}$).  Thus, the Neel vector $\mathbf{l}$ rotates through the angle $\pi$ during one spike, and one can introduce a characteristic time:
\begin{equation}
\Delta t = \pi \alpha_{eff}/\omega_e,
\label{eq:duration}
\end{equation}
which defines the duration of the spike. For $\alpha_{eff}=0.01$ and $\omega_e/2 \pi=1.75\text{GHz}$, which are typical for the AFM  NiO \cite{hutchings1972measurement}, one gets  $\Delta t = 2.85 \text{ps}$.

To achieve the continuous generation of a sequence of spikes, having a ``comb''-like spectrum in the frequency domain, the additional current $j_t$ in the NM layer should be replaced by a periodically modulated control signal:
\begin{equation}
j_e = j_{dc}+j_{ac} \sin 2 \pi f_{ac} t.
\label{eq:sinuscurrent}
\end{equation}
To study the  AFM dynamics under the action of such a control signal, we solved numerically the set of coupled Landau-Lifshitz equations for $\mathbf{M}_1$ and $\mathbf{M}_1$ with the applied current given by Eq.~(\ref{eq:sinuscurrent})(see Eqs. (11-12) in \cite{khymyn2017antiferromagnetic} for details).
This solution was performed in a wide range of the following experimentally controllable parameters of the driving signal and the AFM material: $j_{dc}$ - the amplitude of the dc component of the applied current, $j_{ac}$, $f_{ac}$ - the amplitude and frequency of the ac component of the applied current, and the effective damping constant $\alpha_{eff}$ of the AFM material. In our numerical simulations we assumed a Pt NM layer and the AFM layer made of NiO. We took all the material parameters from Ref.~\cite{khymyn2017antiferromagnetic}, except for the $\alpha_{eff}$.

The examples of the input and output signals of the proposed AFM spike generator are shown on Fig.~\ref{fig:regimes}. Fig.~\ref{fig:regimes} (a) shows the input driving current with a dc component fixed below the threshold $j_{dc}<j_{dc}^{th}$. At the low values of the ac current amplitude $j_{ac}$ the generator does not produce a significant output signal (Fig.~\ref{fig:regimes}(b). With the increase of the $j_{ac}$, at some point when the combined amplitude  exceeds a generation threshold $j_{dc}+j_{ac}=j_1^{th}$, the device generates a single sharp spike of a significant amplitude during each period of applied ac current, (Fig~\ref{fig:regimes} (c), i. e. a periodic sequence of spikes (or a temporal \emph{"comb"}) is generated. For the parameters used in our numerical simulation (see Fig.~\ref{fig:regimes}) the duration of the numerically simulated spike $\Delta t=2.4 \text{ps}$ is close to the value analytically estimated using Eq.(\ref{eq:duration}).  The spike generation has a well-defined threshold on $j_{ac}$, which is similar to the response of a biological neuron to an external stimulus, and the shape of the spike produced by the AFM generator also replicates a typical shape of a spike produced by a biological neuron \cite{CellBiology}. With the further increase of the amplitude of the ac current above the threshold, the Neel vector could switch two ($2\pi$ rotation) or more ($\pi n$ rotation) times per one period of the driving ac current (see Fig.~\ref{fig:regimes} (d). Such a behavior is known for biological neurons as "bursting" \cite{aoyagi2001bursting}. The generation of "bursts"  with a desired number $n$ of spikes, however, requires a fine tuning of the ac amplitude, because the range of the required ac current amplitude $j_{ac}$ rapidly decreases with the increase of $n$.  Thus, below we categorize the reactions of our artificial AFM "neuron" to the external "stimulus" into "no spikes", "single spike", or "bursting" regimes, independently of the number of spikes generated in a "burst" (see Fig.~\ref{fig:diagram}).

The ``phase diagrams'' of the spike generation regimes in an  AFM  auto-oscillator driven by a combined (dc + ac)  control signal are shown on Fig.~\ref{fig:diagram} (a), (b).  Two ``phase diagrams'' are presented: the ``phase diagram'' on the plane  $j_{ac}/ j_{dc}^{th}$ vs. $j_{dc}/ j_{dc}^{th}$  at the fixed values of the driving ac frequency $f_{ac}=$ 20 GHz and the AFM material damping $\alpha_{eff} = 0.01$ (see Fig.~\ref{fig:diagram} (a)), and the ``phase diagram'' on the plane $f_{ac}$ vs. $\alpha_{eff}$ at the fixed values  of  $j_{dc}/ j_{dc}^{th} = 0.8$  and  $j_{ac}/ j_{dc}^{th} = 0.3$ (see Fig.~\ref{fig:diagram} (b)).

As one can see from Fig.~\ref{fig:diagram}(a) the threshold for the spike sequence generation $j_{ac}=j_1^{th}$ lays above the line $j_{ac}+j_{dc}=j_{dc}^{th}$, because the applied ac current, after overcoming the potential barrier caused by the perpendicular anisotropy in the AFM "easy" plane, must produce a sufficient work against the effective damping (see Fig.~\ref{fig:pendulum}). The work produced  by the ac current depends on the duration of its action, and, therefore, on the ac current frequency $f_{ac}$. Consequently, at the fixed value of the dc current $j_{dc}<j_{th}$ the maximum ac current frequency $f_{ac}=f_1^{th}$, at which the generation of the spike sequences  is still possible, decreases with the increase of the effective damping $\alpha_{eff}$  of the AFM material (see Fig.~\ref{fig:diagram} (b)).  We would like to note, that the effective damping $\alpha_{eff} $  can be adjusted not only by choosing a different AFM material having a different intrinsic  damping $\alpha_{0}$,  but also by changing the thickness of the AFM layer, as the spin-pumping-related part of the effective damping is inversely proportional to the AFM later thickness.

Note, that the maximum ``clock'' frequency of spike generation in the  proposed artificial AFM ``neuron'' decreases with the increase of the  effective damping, and is ultimately limited by the inverse duration of a single spike determined by Eq.(\ref{eq:duration}). Our numerical simulations also show that this frequency can be increased up to $f_{ac}>150\text{GHz}$ by the increase of the bias dc current ($j_{dc}\simeq j_{dc}^{th}$) in a  narrow range of the $\alpha_{eff}$ values.

A continuous generation of ultra-short pulses with a random initial phase is also possible in the absence of  driving ac current by applying dc bias current with low supercriticality $0<\left(j_{dc}-j_{dc}^{th}\right)/j_{dc}^{th} \ll 1$, see right bottom corner in Fig.~\ref{fig:diagram}(a). The frequency of the pulse train in this case rapidly increases with the value of dc current\cite{khymyn2017antiferromagnetic} as $f \sim \sqrt{j_{dc}-j_{dc}^{th}}$ and is not stable under the fluctuations of the dc bias. However, such regime of the spikes generation can be phase-locked by the injection of ac control signal, which is indicated by the bending of the single-spike phase for small values of $j_{ac}$ in Fig.~\ref{fig:diagram}(a). This effect to be considered in detail elsewhere.


The application of the proposed AFM generator of ultrashort spikes for traditional signal processing purposes, for example, as a spintronic frequency multiplexer, is also possible, and it requires a sufficiently wide frequency bandwidth of the generated signal.  The simulated spectral density of a spike sequence generated by the above described AFM auto-oscillator at the ac driving frequency of $f_{ac}=15\text{GHz}$ is shown on Fig.~\ref{fig:FrequencyComb}. The spectrum represents a well-known \emph{frequency ``comb''} with a slow decay of the amplitude of higher harmonics with the increase of the harmonic number. The generation bandwidth, which is, obviously, defined by the duration of the single spike, reaches the value of  -- $\Delta f \simeq 200\text{GHz}$  at $-10\text{dB}$. This value can be tuned by the choice of the parameters of a particular AFM auto-oscillator.  In particular, that can be done by tuning the thickness of the AFM layer, and , therefore, tuning the effective damping parameter $\alpha_{eff}$, or by choosing a different AFM materal having a proper value of the in-plane anisotropy (see Eq.~(\ref{eq:duration}).

As a final remark, we would  like to note, that the proposed mechanism of the AFM-based ultra-short spike generation is efficient for \emph{relatively high values of the damping constant} $\alpha_{eff}\geq 0.01$ . This means that metallic AFM materials, like $Mn_{2}Au$  or IrMn could be more suitable for the practical design of the AFM-based  spike generators, than the dielectric AFM , like NiO. The use of conductive AFM layers in a spike generator can also substantially enhance the magnitude of the output signal by employing  the ``AFM tunneling magnetoresistance effect''  \cite{bib:Wadley:2016, bib:Moriyama:2017} instead of the ISHE in the adjacent Pt layer to extract the output spike signal from the AFM material.

\subsection*{Acknowledgements} This work was supported by the Knut and Alice Wallenberg Foundation (KAW). This work was also supported in part by the Grants Nos. EFMA- 1641989 and ECCS-1708982 from the NSF of the USA, and by the DARPA M3IC Grant under the Contract No. W911-17-C-0031. IL acknowledges support from the Russian Science Foundation (project No 14-19-00760).
\subsection*{Correspondence}
Correspondence and requests for materials should be addressed to I.L.\\(email: ivan.lisenkov@phystech.edu).

\bibliography{CombManuscript}

\begin{thebibliography}{24}%
\makeatletter
\providecommand \@ifxundefined [1]{%
 \@ifx{#1\undefined}
}%
\providecommand \@ifnum [1]{%
 \ifnum #1\expandafter \@firstoftwo
 \else \expandafter \@secondoftwo
 \fi
}%
\providecommand \@ifx [1]{%
 \ifx #1\expandafter \@firstoftwo
 \else \expandafter \@secondoftwo
 \fi
}%
\providecommand \natexlab [1]{#1}%
\providecommand \enquote  [1]{``#1''}%
\providecommand \bibnamefont  [1]{#1}%
\providecommand \bibfnamefont [1]{#1}%
\providecommand \citenamefont [1]{#1}%
\providecommand \href@noop [0]{\@secondoftwo}%
\providecommand \href [0]{\begingroup \@sanitize@url \@href}%
\providecommand \@href[1]{\@@startlink{#1}\@@href}%
\providecommand \@@href[1]{\endgroup#1\@@endlink}%
\providecommand \@sanitize@url [0]{\catcode `\\12\catcode `\$12\catcode
  `\&12\catcode `\#12\catcode `\^12\catcode `\_12\catcode `\%12\relax}%
\providecommand \@@startlink[1]{}%
\providecommand \@@endlink[0]{}%
\providecommand \url  [0]{\begingroup\@sanitize@url \@url }%
\providecommand \@url [1]{\endgroup\@href {#1}{\urlprefix }}%
\providecommand \urlprefix  [0]{URL }%
\providecommand \Eprint [0]{\href }%
\providecommand \doibase [0]{http://dx.doi.org/}%
\providecommand \selectlanguage [0]{\@gobble}%
\providecommand \bibinfo  [0]{\@secondoftwo}%
\providecommand \bibfield  [0]{\@secondoftwo}%
\providecommand \translation [1]{[#1]}%
\providecommand \BibitemOpen [0]{}%
\providecommand \bibitemStop [0]{}%
\providecommand \bibitemNoStop [0]{.\EOS\space}%
\providecommand \EOS [0]{\spacefactor3000\relax}%
\providecommand \BibitemShut  [1]{\csname bibitem#1\endcsname}%
\let\auto@bib@innerbib\@empty
\bibitem [{\citenamefont {Berk}\ \emph {et~al.}(1999)\citenamefont {Berk},
  \citenamefont {Zipursky}, \citenamefont {Matsudaira}, \citenamefont {David},\
  and\ \citenamefont {Darnell}}]{CellBiology}%
  \BibitemOpen
  \bibfield  {author} {\bibinfo {author} {\bibfnamefont {A.}~\bibnamefont
  {Berk}}, \bibinfo {author} {\bibfnamefont {S.~L.}\ \bibnamefont {Zipursky}},
  \bibinfo {author} {\bibfnamefont {P.}~\bibnamefont {Matsudaira}}, \bibinfo
  {author} {\bibfnamefont {B.}~\bibnamefont {David}}, \ and\ \bibinfo {author}
  {\bibfnamefont {J.}~\bibnamefont {Darnell}},\ }\href@noop {} {\emph {\bibinfo
  {title} {Molecular Cell Biology}}},\ \bibinfo {edition} {forth}\ ed.\
  (\bibinfo  {publisher} {W H Freeman \& Co (Sd)},\ \bibinfo {year}
  {1999})\BibitemShut {NoStop}%
\bibitem [{\citenamefont {Indiveri}\ \emph {et~al.}(2011)\citenamefont
  {Indiveri}, \citenamefont {Linares-Barranco}, \citenamefont {Hamilton},
  \citenamefont {Van~Schaik}, \citenamefont {Etienne-Cummings}, \citenamefont
  {Delbruck}, \citenamefont {Liu}, \citenamefont {Dudek}, \citenamefont
  {H{\"a}fliger}, \citenamefont {Renaud} \emph
  {et~al.}}]{indiveri2011neuromorphic}%
  \BibitemOpen
  \bibfield  {author} {\bibinfo {author} {\bibfnamefont {G.}~\bibnamefont
  {Indiveri}}, \bibinfo {author} {\bibfnamefont {B.}~\bibnamefont
  {Linares-Barranco}}, \bibinfo {author} {\bibfnamefont {T.~J.}\ \bibnamefont
  {Hamilton}}, \bibinfo {author} {\bibfnamefont {A.}~\bibnamefont
  {Van~Schaik}}, \bibinfo {author} {\bibfnamefont {R.}~\bibnamefont
  {Etienne-Cummings}}, \bibinfo {author} {\bibfnamefont {T.}~\bibnamefont
  {Delbruck}}, \bibinfo {author} {\bibfnamefont {S.-C.}\ \bibnamefont {Liu}},
  \bibinfo {author} {\bibfnamefont {P.}~\bibnamefont {Dudek}}, \bibinfo
  {author} {\bibfnamefont {P.}~\bibnamefont {H{\"a}fliger}}, \bibinfo {author}
  {\bibfnamefont {S.}~\bibnamefont {Renaud}},  \emph {et~al.},\ }\href@noop {}
  {\bibfield  {journal} {\bibinfo  {journal} {Frontiers in neuroscience}\
  }\textbf {\bibinfo {volume} {5}} (\bibinfo {year} {2011})}\BibitemShut
  {NoStop}%
\bibitem [{\citenamefont {Aoyagi}\ \emph {et~al.}(2001)\citenamefont {Aoyagi},
  \citenamefont {Terada}, \citenamefont {Kang}, \citenamefont {Kaneko},\ and\
  \citenamefont {Fukai}}]{aoyagi2001bursting}%
  \BibitemOpen
  \bibfield  {author} {\bibinfo {author} {\bibfnamefont {T.}~\bibnamefont
  {Aoyagi}}, \bibinfo {author} {\bibfnamefont {N.}~\bibnamefont {Terada}},
  \bibinfo {author} {\bibfnamefont {Y.}~\bibnamefont {Kang}}, \bibinfo {author}
  {\bibfnamefont {T.}~\bibnamefont {Kaneko}}, \ and\ \bibinfo {author}
  {\bibfnamefont {T.}~\bibnamefont {Fukai}},\ }\href@noop {} {\bibfield
  {journal} {\bibinfo  {journal} {Neurocomputing}\ }\textbf {\bibinfo {volume}
  {38}},\ \bibinfo {pages} {93} (\bibinfo {year} {2001})}\BibitemShut {NoStop}%
\bibitem [{\citenamefont {Tomita}\ \emph {et~al.}(2013)\citenamefont {Tomita},
  \citenamefont {Suzuki}, \citenamefont {Matsumoto}, \citenamefont {Yamamoto},
  \citenamefont {Yamaguchi},\ and\ \citenamefont {Tamura}}]{tomita20138}%
  \BibitemOpen
  \bibfield  {author} {\bibinfo {author} {\bibfnamefont {Y.}~\bibnamefont
  {Tomita}}, \bibinfo {author} {\bibfnamefont {K.}~\bibnamefont {Suzuki}},
  \bibinfo {author} {\bibfnamefont {T.}~\bibnamefont {Matsumoto}}, \bibinfo
  {author} {\bibfnamefont {T.}~\bibnamefont {Yamamoto}}, \bibinfo {author}
  {\bibfnamefont {H.}~\bibnamefont {Yamaguchi}}, \ and\ \bibinfo {author}
  {\bibfnamefont {H.}~\bibnamefont {Tamura}},\ }in\ \href@noop {} {\emph
  {\bibinfo {booktitle} {VLSI Circuits (VLSIC), 2013 Symposium on}}}\ (\bibinfo
  {organization} {IEEE},\ \bibinfo {year} {2013})\ pp.\ \bibinfo {pages}
  {C238--C239}\BibitemShut {NoStop}%
\bibitem [{\citenamefont {Saitoh}\ \emph {et~al.}(1995)\citenamefont {Saitoh},
  \citenamefont {Kourogi},\ and\ \citenamefont {Ohtsu}}]{saitoh1995waveguide}%
  \BibitemOpen
  \bibfield  {author} {\bibinfo {author} {\bibfnamefont {T.}~\bibnamefont
  {Saitoh}}, \bibinfo {author} {\bibfnamefont {M.}~\bibnamefont {Kourogi}}, \
  and\ \bibinfo {author} {\bibfnamefont {M.}~\bibnamefont {Ohtsu}},\
  }\href@noop {} {\bibfield  {journal} {\bibinfo  {journal} {IEEE photonics
  technology letters}\ }\textbf {\bibinfo {volume} {7}},\ \bibinfo {pages}
  {197} (\bibinfo {year} {1995})}\BibitemShut {NoStop}%
\bibitem [{\citenamefont {Park}\ and\ \citenamefont
  {Wysocki}(1996)}]{park199624}%
  \BibitemOpen
  \bibfield  {author} {\bibinfo {author} {\bibfnamefont {N.}~\bibnamefont
  {Park}}\ and\ \bibinfo {author} {\bibfnamefont {P.~F.}\ \bibnamefont
  {Wysocki}},\ }\href@noop {} {\bibfield  {journal} {\bibinfo  {journal} {IEEE
  Photonics Technology Letters}\ }\textbf {\bibinfo {volume} {8}},\ \bibinfo
  {pages} {1459} (\bibinfo {year} {1996})}\BibitemShut {NoStop}%
\bibitem [{\citenamefont {Yamashita}\ and\ \citenamefont
  {Cowle}(1998)}]{yamashita1998bidirectional}%
  \BibitemOpen
  \bibfield  {author} {\bibinfo {author} {\bibfnamefont {S.}~\bibnamefont
  {Yamashita}}\ and\ \bibinfo {author} {\bibfnamefont {G.~J.}\ \bibnamefont
  {Cowle}},\ }\href@noop {} {\bibfield  {journal} {\bibinfo  {journal} {IEEE
  Photonics Technology Letters}\ }\textbf {\bibinfo {volume} {10}},\ \bibinfo
  {pages} {796} (\bibinfo {year} {1998})}\BibitemShut {NoStop}%
\bibitem [{\citenamefont {Bennett}\ \emph {et~al.}(1999)\citenamefont
  {Bennett}, \citenamefont {Cai}, \citenamefont {Burr}, \citenamefont {Gough},\
  and\ \citenamefont {Seeds}}]{bennett19991}%
  \BibitemOpen
  \bibfield  {author} {\bibinfo {author} {\bibfnamefont {S.}~\bibnamefont
  {Bennett}}, \bibinfo {author} {\bibfnamefont {B.}~\bibnamefont {Cai}},
  \bibinfo {author} {\bibfnamefont {E.}~\bibnamefont {Burr}}, \bibinfo {author}
  {\bibfnamefont {O.}~\bibnamefont {Gough}}, \ and\ \bibinfo {author}
  {\bibfnamefont {A.}~\bibnamefont {Seeds}},\ }\href@noop {} {\bibfield
  {journal} {\bibinfo  {journal} {IEEE Photonics Technology Letters}\ }\textbf
  {\bibinfo {volume} {11}},\ \bibinfo {pages} {551} (\bibinfo {year}
  {1999})}\BibitemShut {NoStop}%
\bibitem [{\citenamefont {Kiselev}\ \emph {et~al.}(2003)\citenamefont
  {Kiselev}, \citenamefont {Sankey}, \citenamefont {Krivorotov}, \citenamefont
  {Emley}, \citenamefont {Schoelkopf}, \citenamefont {Buhrman},\ and\
  \citenamefont {Ralph}}]{Kiselev2003}%
  \BibitemOpen
  \bibfield  {author} {\bibinfo {author} {\bibfnamefont {S.~I.}\ \bibnamefont
  {Kiselev}}, \bibinfo {author} {\bibfnamefont {J.~C.}\ \bibnamefont {Sankey}},
  \bibinfo {author} {\bibfnamefont {I.~N.}\ \bibnamefont {Krivorotov}},
  \bibinfo {author} {\bibfnamefont {N.~C.}\ \bibnamefont {Emley}}, \bibinfo
  {author} {\bibfnamefont {R.~J.}\ \bibnamefont {Schoelkopf}}, \bibinfo
  {author} {\bibfnamefont {R.~A.}\ \bibnamefont {Buhrman}}, \ and\ \bibinfo
  {author} {\bibfnamefont {D.~C.}\ \bibnamefont {Ralph}},\ }\href {\doibase
  10.1038/nature01967} {\bibfield  {journal} {\bibinfo  {journal} {Nature}\
  }\textbf {\bibinfo {volume} {425}},\ \bibinfo {pages} {380} (\bibinfo {year}
  {2003})}\BibitemShut {NoStop}%
\bibitem [{\citenamefont {Demidov}\ \emph {et~al.}(2012)\citenamefont
  {Demidov}, \citenamefont {Urazhdin}, \citenamefont {Ulrichs}, \citenamefont
  {Tiberkevich}, \citenamefont {Slavin}, \citenamefont {Baither}, \citenamefont
  {Schmitz},\ and\ \citenamefont {Demokritov}}]{Demidov2012}%
  \BibitemOpen
  \bibfield  {author} {\bibinfo {author} {\bibfnamefont {V.~E.}\ \bibnamefont
  {Demidov}}, \bibinfo {author} {\bibfnamefont {S.}~\bibnamefont {Urazhdin}},
  \bibinfo {author} {\bibfnamefont {H.}~\bibnamefont {Ulrichs}}, \bibinfo
  {author} {\bibfnamefont {V.}~\bibnamefont {Tiberkevich}}, \bibinfo {author}
  {\bibfnamefont {A.}~\bibnamefont {Slavin}}, \bibinfo {author} {\bibfnamefont
  {D.}~\bibnamefont {Baither}}, \bibinfo {author} {\bibfnamefont
  {G.}~\bibnamefont {Schmitz}}, \ and\ \bibinfo {author} {\bibfnamefont
  {S.~O.}\ \bibnamefont {Demokritov}},\ }\href {\doibase 10.1038/nmat3459}
  {\bibfield  {journal} {\bibinfo  {journal} {Nature Materials}\ }\textbf
  {\bibinfo {volume} {11}},\ \bibinfo {pages} {1028–1031} (\bibinfo {year}
  {2012})}\BibitemShut {NoStop}%
\bibitem [{\citenamefont {Collet}\ \emph {et~al.}(2016)\citenamefont {Collet},
  \citenamefont {de~Milly}, \citenamefont {d'~Allivy~Kelly}, \citenamefont
  {Naletov}, \citenamefont {Bernard}, \citenamefont {Bortolotti}, \citenamefont
  {Ben~Youssef}, \citenamefont {Demidov}, \citenamefont {Demokritov},
  \citenamefont {Prieto},\ and\ \citenamefont {et~al.}}]{Demidov2016}%
  \BibitemOpen
  \bibfield  {author} {\bibinfo {author} {\bibfnamefont {M.}~\bibnamefont
  {Collet}}, \bibinfo {author} {\bibfnamefont {X.}~\bibnamefont {de~Milly}},
  \bibinfo {author} {\bibfnamefont {O.}~\bibnamefont {d'~Allivy~Kelly}},
  \bibinfo {author} {\bibfnamefont {V.~V.}\ \bibnamefont {Naletov}}, \bibinfo
  {author} {\bibfnamefont {R.}~\bibnamefont {Bernard}}, \bibinfo {author}
  {\bibfnamefont {P.}~\bibnamefont {Bortolotti}}, \bibinfo {author}
  {\bibfnamefont {J.}~\bibnamefont {Ben~Youssef}}, \bibinfo {author}
  {\bibfnamefont {V.~E.}\ \bibnamefont {Demidov}}, \bibinfo {author}
  {\bibfnamefont {S.~O.}\ \bibnamefont {Demokritov}}, \bibinfo {author}
  {\bibfnamefont {J.~L.}\ \bibnamefont {Prieto}}, \ and\ \bibinfo {author}
  {\bibnamefont {et~al.}},\ }\href {\doibase 10.1038/ncomms10377} {\bibfield
  {journal} {\bibinfo  {journal} {Nat. Commun.}\ }\textbf {\bibinfo {volume}
  {7}},\ \bibinfo {pages} {10377} (\bibinfo {year} {2016})}\BibitemShut
  {NoStop}%
\bibitem [{\citenamefont {Chen}\ \emph {et~al.}(2016)\citenamefont {Chen},
  \citenamefont {Dumas}, \citenamefont {Eklund}, \citenamefont {Muduli},
  \citenamefont {Houshang}, \citenamefont {Awad}, \citenamefont
  {D{\"u}rrenfeld}, \citenamefont {Malm}, \citenamefont {Rusu},\ and\
  \citenamefont {{\AA}kerman}}]{chen2016spin}%
  \BibitemOpen
  \bibfield  {author} {\bibinfo {author} {\bibfnamefont {T.}~\bibnamefont
  {Chen}}, \bibinfo {author} {\bibfnamefont {R.~K.}\ \bibnamefont {Dumas}},
  \bibinfo {author} {\bibfnamefont {A.}~\bibnamefont {Eklund}}, \bibinfo
  {author} {\bibfnamefont {P.~K.}\ \bibnamefont {Muduli}}, \bibinfo {author}
  {\bibfnamefont {A.}~\bibnamefont {Houshang}}, \bibinfo {author}
  {\bibfnamefont {A.~A.}\ \bibnamefont {Awad}}, \bibinfo {author}
  {\bibfnamefont {P.}~\bibnamefont {D{\"u}rrenfeld}}, \bibinfo {author}
  {\bibfnamefont {B.~G.}\ \bibnamefont {Malm}}, \bibinfo {author}
  {\bibfnamefont {A.}~\bibnamefont {Rusu}}, \ and\ \bibinfo {author}
  {\bibfnamefont {J.}~\bibnamefont {{\AA}kerman}},\ }\href@noop {} {\bibfield
  {journal} {\bibinfo  {journal} {Proceedings of the IEEE}\ }\textbf {\bibinfo
  {volume} {104}},\ \bibinfo {pages} {1919} (\bibinfo {year}
  {2016})}\BibitemShut {NoStop}%
\bibitem [{\citenamefont {Mohseni}\ \emph {et~al.}(2013)\citenamefont
  {Mohseni}, \citenamefont {Sani}, \citenamefont {Persson}, \citenamefont
  {Nguyen}, \citenamefont {Chung}, \citenamefont {Pogoryelov}, \citenamefont
  {Muduli}, \citenamefont {Iacocca}, \citenamefont {Eklund}, \citenamefont
  {Dumas} \emph {et~al.}}]{mohseni2013spin}%
  \BibitemOpen
  \bibfield  {author} {\bibinfo {author} {\bibfnamefont {S.~M.}\ \bibnamefont
  {Mohseni}}, \bibinfo {author} {\bibfnamefont {S.}~\bibnamefont {Sani}},
  \bibinfo {author} {\bibfnamefont {J.}~\bibnamefont {Persson}}, \bibinfo
  {author} {\bibfnamefont {T.~A.}\ \bibnamefont {Nguyen}}, \bibinfo {author}
  {\bibfnamefont {S.}~\bibnamefont {Chung}}, \bibinfo {author} {\bibfnamefont
  {Y.}~\bibnamefont {Pogoryelov}}, \bibinfo {author} {\bibfnamefont
  {P.}~\bibnamefont {Muduli}}, \bibinfo {author} {\bibfnamefont
  {E.}~\bibnamefont {Iacocca}}, \bibinfo {author} {\bibfnamefont
  {A.}~\bibnamefont {Eklund}}, \bibinfo {author} {\bibfnamefont
  {R.}~\bibnamefont {Dumas}},  \emph {et~al.},\ }\href@noop {} {\bibfield
  {journal} {\bibinfo  {journal} {Science}\ }\textbf {\bibinfo {volume}
  {339}},\ \bibinfo {pages} {1295} (\bibinfo {year} {2013})}\BibitemShut
  {NoStop}%
\bibitem [{\citenamefont {Gomonay}\ and\ \citenamefont
  {Loktev}(2014)}]{bib:Gomonay:2014}%
  \BibitemOpen
  \bibfield  {author} {\bibinfo {author} {\bibfnamefont {E.~V.}\ \bibnamefont
  {Gomonay}}\ and\ \bibinfo {author} {\bibfnamefont {V.~M.}\ \bibnamefont
  {Loktev}},\ }\href {\doibase 10.1063/1.4862467} {\bibfield  {journal}
  {\bibinfo  {journal} {Low Temp. Phys.}\ }\textbf {\bibinfo {volume} {40}},\
  \bibinfo {pages} {17} (\bibinfo {year} {2014})}\BibitemShut {NoStop}%
\bibitem [{\citenamefont {Cheng}\ \emph {et~al.}(2016)\citenamefont {Cheng},
  \citenamefont {Xiao},\ and\ \citenamefont {Brataas}}]{bib:Cheng:2016}%
  \BibitemOpen
  \bibfield  {author} {\bibinfo {author} {\bibfnamefont {R.}~\bibnamefont
  {Cheng}}, \bibinfo {author} {\bibfnamefont {D.}~\bibnamefont {Xiao}}, \ and\
  \bibinfo {author} {\bibfnamefont {A.}~\bibnamefont {Brataas}},\ }\href
  {\doibase 10.1103/physrevlett.116.207603} {\bibfield  {journal} {\bibinfo
  {journal} {Physical Review Letters}\ }\textbf {\bibinfo {volume} {116}},\
  \bibinfo {pages} {207603} (\bibinfo {year} {2016})}\BibitemShut {NoStop}%
\bibitem [{\citenamefont {Cheng}\ \emph {et~al.}(2014)\citenamefont {Cheng},
  \citenamefont {Xiao}, \citenamefont {Niu},\ and\ \citenamefont
  {Brataas}}]{cheng2014spin}%
  \BibitemOpen
  \bibfield  {author} {\bibinfo {author} {\bibfnamefont {R.}~\bibnamefont
  {Cheng}}, \bibinfo {author} {\bibfnamefont {J.}~\bibnamefont {Xiao}},
  \bibinfo {author} {\bibfnamefont {Q.}~\bibnamefont {Niu}}, \ and\ \bibinfo
  {author} {\bibfnamefont {A.}~\bibnamefont {Brataas}},\ }\href@noop {}
  {\bibfield  {journal} {\bibinfo  {journal} {Physical review letters}\
  }\textbf {\bibinfo {volume} {113}},\ \bibinfo {pages} {057601} (\bibinfo
  {year} {2014})}\BibitemShut {NoStop}%
\bibitem [{\citenamefont {Liu}\ \emph {et~al.}(2016)\citenamefont {Liu},
  \citenamefont {Yin}, \citenamefont {Zang}, \citenamefont {Lake},\ and\
  \citenamefont {Barlas}}]{bib:Liu:2016}%
  \BibitemOpen
  \bibfield  {author} {\bibinfo {author} {\bibfnamefont {Y.}~\bibnamefont
  {Liu}}, \bibinfo {author} {\bibfnamefont {G.}~\bibnamefont {Yin}}, \bibinfo
  {author} {\bibfnamefont {J.}~\bibnamefont {Zang}}, \bibinfo {author}
  {\bibfnamefont {R.~K.}\ \bibnamefont {Lake}}, \ and\ \bibinfo {author}
  {\bibfnamefont {Y.}~\bibnamefont {Barlas}},\ }\href {\doibase
  10.1103/physrevb.94.094434} {\bibfield  {journal} {\bibinfo  {journal}
  {Physical Review B}\ }\textbf {\bibinfo {volume} {94}} (\bibinfo {year}
  {2016}),\ 10.1103/physrevb.94.094434}\BibitemShut {NoStop}%
\bibitem [{\citenamefont {Khymyn}\ \emph {et~al.}(2017)\citenamefont {Khymyn},
  \citenamefont {Lisenkov}, \citenamefont {Tiberkevich}, \citenamefont
  {Ivanov},\ and\ \citenamefont {Slavin}}]{khymyn2017antiferromagnetic}%
  \BibitemOpen
  \bibfield  {author} {\bibinfo {author} {\bibfnamefont {R.}~\bibnamefont
  {Khymyn}}, \bibinfo {author} {\bibfnamefont {I.}~\bibnamefont {Lisenkov}},
  \bibinfo {author} {\bibfnamefont {V.}~\bibnamefont {Tiberkevich}}, \bibinfo
  {author} {\bibfnamefont {B.~A.}\ \bibnamefont {Ivanov}}, \ and\ \bibinfo
  {author} {\bibfnamefont {A.}~\bibnamefont {Slavin}},\ }\href@noop {}
  {\bibfield  {journal} {\bibinfo  {journal} {Scientific Reports}\ }\textbf
  {\bibinfo {volume} {7}} (\bibinfo {year} {2017})}\BibitemShut {NoStop}%
\bibitem [{\citenamefont {Sulymenko}\ \emph {et~al.}(2017)\citenamefont
  {Sulymenko}, \citenamefont {Prokopenko}, \citenamefont {Tiberkevich},
  \citenamefont {Slavin}, \citenamefont {Ivanov},\ and\ \citenamefont
  {Khymyn}}]{sulymenko2017terahertz}%
  \BibitemOpen
  \bibfield  {author} {\bibinfo {author} {\bibfnamefont {O.}~\bibnamefont
  {Sulymenko}}, \bibinfo {author} {\bibfnamefont {O.}~\bibnamefont
  {Prokopenko}}, \bibinfo {author} {\bibfnamefont {V.}~\bibnamefont
  {Tiberkevich}}, \bibinfo {author} {\bibfnamefont {A.}~\bibnamefont {Slavin}},
  \bibinfo {author} {\bibfnamefont {B.}~\bibnamefont {Ivanov}}, \ and\ \bibinfo
  {author} {\bibfnamefont {R.}~\bibnamefont {Khymyn}},\ }\href@noop {}
  {\bibfield  {journal} {\bibinfo  {journal} {Physical Review Applied}\
  }\textbf {\bibinfo {volume} {8}},\ \bibinfo {pages} {064007} (\bibinfo {year}
  {2017})}\BibitemShut {NoStop}%
\bibitem [{\citenamefont {Tserkovnyak}\ \emph {et~al.}(2002)\citenamefont
  {Tserkovnyak}, \citenamefont {Brataas},\ and\ \citenamefont
  {Bauer}}]{Tserkovnyak2002}%
  \BibitemOpen
  \bibfield  {author} {\bibinfo {author} {\bibfnamefont {Y.}~\bibnamefont
  {Tserkovnyak}}, \bibinfo {author} {\bibfnamefont {A.}~\bibnamefont
  {Brataas}}, \ and\ \bibinfo {author} {\bibfnamefont {G.~E.~W.}\ \bibnamefont
  {Bauer}},\ }\href {\doibase 10.1103/physrevlett.88.117601} {\bibfield
  {journal} {\bibinfo  {journal} {Physical Review Letters}\ }\textbf {\bibinfo
  {volume} {88}} (\bibinfo {year} {2002}),\
  10.1103/physrevlett.88.117601}\BibitemShut {NoStop}%
\bibitem [{\citenamefont {Nakayama}\ \emph {et~al.}(2012)\citenamefont
  {Nakayama}, \citenamefont {Ando}, \citenamefont {Harii}, \citenamefont
  {Yoshino}, \citenamefont {Takahashi}, \citenamefont {Kajiwara}, \citenamefont
  {Uchida}, \citenamefont {Fujikawa},\ and\ \citenamefont
  {Saitoh}}]{bib:Nakayama:2012}%
  \BibitemOpen
  \bibfield  {author} {\bibinfo {author} {\bibfnamefont {H.}~\bibnamefont
  {Nakayama}}, \bibinfo {author} {\bibfnamefont {K.}~\bibnamefont {Ando}},
  \bibinfo {author} {\bibfnamefont {K.}~\bibnamefont {Harii}}, \bibinfo
  {author} {\bibfnamefont {T.}~\bibnamefont {Yoshino}}, \bibinfo {author}
  {\bibfnamefont {R.}~\bibnamefont {Takahashi}}, \bibinfo {author}
  {\bibfnamefont {Y.}~\bibnamefont {Kajiwara}}, \bibinfo {author}
  {\bibfnamefont {K.}~\bibnamefont {Uchida}}, \bibinfo {author} {\bibfnamefont
  {Y.}~\bibnamefont {Fujikawa}}, \ and\ \bibinfo {author} {\bibfnamefont
  {E.}~\bibnamefont {Saitoh}},\ }\href {\doibase 10.1103/physrevb.85.144408}
  {\bibfield  {journal} {\bibinfo  {journal} {Physical Review B}\ }\textbf
  {\bibinfo {volume} {85}},\ \bibinfo {pages} {144408} (\bibinfo {year}
  {2012})}\BibitemShut {NoStop}%
\bibitem [{\citenamefont {Hutchings}\ and\ \citenamefont
  {Samuelsen}(1972)}]{hutchings1972measurement}%
  \BibitemOpen
  \bibfield  {author} {\bibinfo {author} {\bibfnamefont {M.~T.}\ \bibnamefont
  {Hutchings}}\ and\ \bibinfo {author} {\bibfnamefont {E.}~\bibnamefont
  {Samuelsen}},\ }\href@noop {} {\bibfield  {journal} {\bibinfo  {journal}
  {Physical Review B}\ }\textbf {\bibinfo {volume} {6}},\ \bibinfo {pages}
  {3447} (\bibinfo {year} {1972})}\BibitemShut {NoStop}%
\bibitem [{\citenamefont {Wadley}\ \emph {et~al.}(2016)\citenamefont {Wadley},
  \citenamefont {Howells}, \citenamefont { elezny}, \citenamefont {Andrews},
  \citenamefont {Hills}, \citenamefont {Campion}, \citenamefont {Novak},
  \citenamefont {Olejnik}, \citenamefont {Maccherozzi}, \citenamefont {Dhesi},\
  and\ \citenamefont {et~al.}}]{bib:Wadley:2016}%
  \BibitemOpen
  \bibfield  {author} {\bibinfo {author} {\bibfnamefont {P.}~\bibnamefont
  {Wadley}}, \bibinfo {author} {\bibfnamefont {B.}~\bibnamefont {Howells}},
  \bibinfo {author} {\bibfnamefont {J.}~\bibnamefont { elezny}}, \bibinfo
  {author} {\bibfnamefont {C.}~\bibnamefont {Andrews}}, \bibinfo {author}
  {\bibfnamefont {V.}~\bibnamefont {Hills}}, \bibinfo {author} {\bibfnamefont
  {R.~P.}\ \bibnamefont {Campion}}, \bibinfo {author} {\bibfnamefont
  {V.}~\bibnamefont {Novak}}, \bibinfo {author} {\bibfnamefont
  {K.}~\bibnamefont {Olejnik}}, \bibinfo {author} {\bibfnamefont
  {F.}~\bibnamefont {Maccherozzi}}, \bibinfo {author} {\bibfnamefont {S.~S.}\
  \bibnamefont {Dhesi}}, \ and\ \bibinfo {author} {\bibnamefont {et~al.}},\
  }\href {\doibase 10.1126/science.aab1031} {\bibfield  {journal} {\bibinfo
  {journal} {Science}\ }\textbf {\bibinfo {volume} {351}},\ \bibinfo {pages}
  {587–590} (\bibinfo {year} {2016})}\BibitemShut {NoStop}%
\bibitem [{\citenamefont {Moriyama}\ \emph {et~al.}(2017)\citenamefont
  {Moriyama}, \citenamefont {Oda},\ and\ \citenamefont
  {Ono}}]{bib:Moriyama:2017}%
  \BibitemOpen
  \bibfield  {author} {\bibinfo {author} {\bibfnamefont {T.}~\bibnamefont
  {Moriyama}}, \bibinfo {author} {\bibfnamefont {K.}~\bibnamefont {Oda}}, \
  and\ \bibinfo {author} {\bibfnamefont {T.}~\bibnamefont {Ono}},\ }\href@noop
  {} {\enquote {\bibinfo {title} {Spin torque control of antiferromagnetic
  moments in nio},}\ } (\bibinfo {year} {2017}),\ \Eprint
  {http://arxiv.org/abs/1708.07682} {arXiv:1708.07682} \BibitemShut {NoStop}%
\end{thebibliography}%

\end{document}